\RequirePackage{fix-cm}

\documentclass[smallextended]{svjour3}       

\smartqed  

\usepackage{graphicx}
\usepackage{mathptmx}      
\usepackage{amsmath}
\usepackage{latexsym}

\begin{document}

\title{K-essential covariant holography
}

\author{Alberto Rozas-Fern\'andez
}

\institute{Alberto Rozas-Fern\'andez \at
              Instituto de F\'{\i}sica Fundamental,
Consejo Superior de Investigaciones Cient\'{\i}ficas, Serrano 121,
28006 Madrid, Spain
\and
Institute of Cosmology and Gravitation, University of Portsmouth, Dennis Sciama Building, Portsmouth PO1 3FX, United Kingdom.
              \email{a.rozas@iff.csic.es}
}

\date{Received: date / Accepted: date}

\maketitle

\begin{abstract}
The holographic principle is applied to a flat Friedmann-Robertson-Walker space-time dominated by dark energy when this is due to the presence of a k-essence scalar field, both for dark energy and phantom scenarios. In this framework, a geometrical covariant approach permits the construction of holographic hypersurfaces. The resulting covariant preferred screens, both for phantom and non-phantom regions, are then compared with those obtained by using the holographic dark energy model with the future event horizon as the infrared cut-off.  In the phantom case, one of the two obtained holographic screens is placed on the big rip hypersurface, both for the covariant holographic formalism and the holographic phantom model. It is also analysed whether this covariant formalism allows a mathematically consistent formulation of fundamental theories based on the existence of a S-matrix at infinite distances.
\end{abstract}

\section{Introduction}

The holographic principle \cite{'tHooft:1993gx,Susskind:1994vu} is a relation between spacetime geometry and the number of degrees of freedom and implies that all information contained on a light-like hypersurface $L$ can be stored on the boundary of it, $B$, at a density of no more than one bit per Planck area. The most explicit example of the holographic principle is the AdS/CFT correspondence \cite{Maldacena:1997re}. On the other hand, a proposal for a holographic description of cosmology was made in \cite{Fischler:1998st}. See \cite{Bousso:2002ju} for a review on the holographic principle.

Our aim is this work is to apply the holographic principle to the spacetimes of accelerating universes filled with a k-essence scalar field with a constant equation of state (EoS) $w=p/\rho$, both for the region $-1<w<-1/3$ and the phantom region $w<-1$. We shall do so by means of a covariant procedure \cite{Bousso:2002ju} which will be then compared with the results obtained using the holographic dark energy (HDE) model with the future event horizon as the infrared cutoff \cite{Li:2004rb}. The interest of our analysis is two-fold. On the one hand, dark energy should contain a large amount of the relevant degrees of freedom and hence, in order to constrain the EoS for dark energy, it is important to investigate whether such degrees of freedom are projected on the same boundary surfaces as those characterising the remaining non-vacuum energy. On the other hand, a universe with constant EoS $w$ that accelerates indefinitely will exhibit a future event horizon \cite{He:2001za} (see, however, \cite{GonzalezDiaz:2001ce}), presenting a challenge for string theories because it is not possible to construct a conventional S-matrix as the local observer inside his horizon is not able to isolate particles to be scattered. In the HDE model the future event horizon, which would behave as a holographic screen, would exacerbate this problem.

The rest of the paper can be outlined as follows. In Sec.\ \ref{sec:kessenceDE} we introduce the spacetime generated by a flat Friedmann-Robertson-Walker (FRW) k-essence dominated universe in the region $w>-1$. In Sec.\ \ref{sec:covkessenceDE} we use a covariant formalism to
derive the ultimate holographic preferred screens that correspond to
the spacetime shown in Sec.\ \ref{sec:kessenceDE}. In Sec.\ \ref{sec:phkessenceDE} we discuss
the covariant holography of a flat FRW k-essential phantom energy scenario. In Sec.\ \ref{sec:HDE}, we construct
the dark and phantom HDE models for a flat geometry in order to insert
holographic screens in terms of the future event
horizon \cite{Li:2004rb} or the horizon at the big rip \cite{GonzalezDiaz:2005sh}. The conclusions are drawn in Sec.\ \ref{sec:concl}.

\section{A k-essential dark energy universe}\label{sec:kessenceDE}

Understanding the nature of dark energy is arguably the most important challenge in modern cosmology. If we consider that dark energy is dynamical \cite{Samushia:2012iq} and is due to the presence of a
k-essence scalar field \cite{Chiba:1999ka,ArmendarizPicon:2000dh,ArmendarizPicon:2000ah}, this will give rise to a
variety of spacetime structures for the universe whose holographic properties deserve to be studied.

As a dark energy candidate, k-essence is in general defined as a scalar field $\phi$ with non-canonical kinetic
energy associated with a Lagrangian
\begin{equation}
L= K(\phi)p(X),
\end{equation} where $K(\phi)>0$ and $X=\frac{1}{2}\nabla_{\mu}\phi\nabla^{\mu}\phi$ is the kinetic energy of the field $\phi$. In
the case of the k-essence scalar field, the negative pressure
that explains the accelerated expansion arises out of
modifications to the kinetic energy $X$. Quintessence and
tachyonic models belong to k-essence. In k-essence, the
higher order terms are not necessarily negligible which,
interestingly, can give rise to new dynamics not possible
in quintessence. Every quintessence model can be
viewed as a k-essence model generated by a kinetic linear
function. On the other hand, the tachyon model is
classified as k-essence because it belongs to a class of
the action for the k-essence. However, in the sense that
the kinetic energy of the tachyon needs to be suppressed
to realise cosmic acceleration, this scenario is different
from k-essence.

 For convenience we choose $K(\phi)=1/\phi^{2}$ \cite{ArmendarizPicon:2000dh,ArmendarizPicon:2000ah} and define the quantities $g(y) \equiv p(X)y$ and $y \equiv 1/\sqrt{X}$.

Using the perfect fluid analogy and for this choice of $K(\phi)$, the energy density and pressure are given by
\begin{equation}
\rho_{\phi} = -g'(y)/\phi^{2}\;,
\end{equation}

\begin{equation}
p_{\phi} = \frac{g(y)}{y\phi^{2}}\;,
\end{equation}
where $g'(y) \equiv dg/dy$.

Assuming that the weak energy condition holds, $\rho_{\phi}>0$, it follows that $g'(y)<0$.

The EoS $w$ and
the effective sound speed of k-essence are given by
\begin{equation}
w =\frac{p_{\phi}}{\rho_{\phi}}= -\frac{g(y)}{yg'(y)}\;,
\end{equation}

\begin{equation}\label{sofs}
c_{s}^{2} =\frac{p_{\phi}'}{\rho_{\phi}'}= \frac{g(y)-yg'(y)}{y^2g''(y)}\;.
\end{equation}

The existence of the late-time accelerated solution requires the condition $w<-1/3$, which implies that $g(y)<0$. We also assume that $w>-1$ and $c_{s}^{2}>0$ and hence $g''(y)>0$. As a
result, $g(y)$ must be a convex and decreasing function of $y$ \cite{ArmendarizPicon:2000dh,ArmendarizPicon:2000ah}.

A most simple family of g-functions satisfying the above requirements is
\begin{equation}\label{g}
g(y)=Ay^{\alpha},
\end{equation} where $A$ and $\alpha$ are given constants such that $A<0$ and $\alpha \left(=-\frac{1}{w}\right)$ satisfies the observational constraint $0<\alpha<0.9$ \cite{Ade:2013zuv}. As a matter of fact, a more general function $g(y)$ can be expressed in terms of polynomials
\begin{equation}\label{gpol}
g(y)=\sum_{i}A_{i}y^{\alpha_{i}},
\end{equation} where the first term is given by Eq. (\ref{g}) and the other extra terms have powers $0<\alpha_{i}<0.9$ and coefficients $A<A_{1}<A_{2}<...$. Nevertheless, for the aims of our study, it will suffice to take only the first term of Eq. (\ref{gpol}).

\begin{figure}[tbp]
\begin{center}
\includegraphics[width=.5\columnwidth]{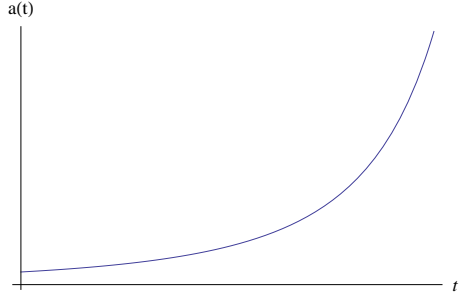}
\caption{\label{fig:1}Generic shape of the evolution of the scale factor $a(t)$ with cosmological time $t$ for the function $g(y)$ given in Eq. (\ref{g}). All units in the plot are arbitrary.}
\end{center}
\end{figure}

In what follows, we shall consider a spatially flat FRW spacetime with line element
\begin{equation}\label{metric}
ds^{2}=-dt^{2}+a(t)^{2}d\vec{x}^{2},
\end{equation} in which $a(t)$ is the scale factor. In the case of a universe dominated by a k-essence scalar field, the Einstein field equations are given by

\begin{equation}\label{Einsteineqns1}
3H^{2}=\rho_{\phi}
\end{equation}

\begin{equation}\label{Einsteineqns2}
2\dot{H}+\rho_{\phi}+p_{\phi}=0,
\end{equation} where $H=\dot{a}/a$ and the overhead dot is the derivative with respect to cosmic time.
If we now combine Eqs. (\ref{Einsteineqns1}), (\ref{Einsteineqns2}) and (\ref{g}), we arrive at
\begin{equation}
3H^{2}=\frac{2\dot{H}\alpha}{1-\alpha}.
\end{equation}
Therefore, the scale factor is expressed as
\begin{equation}\label{scalefactor}
a(t)=\left[a_{0}^{-\frac{3(1-\alpha)}{2\alpha}}-\frac{3}{2}\left(\frac{1-\alpha}{\alpha}\right)(t-t_{0})\right]^{-2\alpha/[3(1-\alpha)]}.
\end{equation} The evolution of $a(t)$ is depicted in Fig.\ \ref{fig:1}.

In order to investigate the holographic properties of the
considered spacetime we express solution
(\ref{scalefactor}) in terms of the conformal time for convenience

\begin{equation}\label{conftime}
\eta=\int\frac{dt}{a(t)}=\frac{2\alpha}{(\alpha-3)}a^{-(3-\alpha)/2\alpha}.
\end{equation}
Note that $0 <\eta <\infty$ for $w > -1/3$ (i.e. $\alpha>3$), and $-\infty <\eta
< 0 $ for $w < -1/3$ (i.e. $\alpha<3$).

Inserting Eq. (\ref{conftime}) in Eq. (\ref{scalefactor}) yields the scale factor as a function of the conformal time
\begin{equation}\label{scalefactorconf}
a(\eta)=\left[\frac{(\alpha-3)\eta}{2\alpha}\right]^{-2\alpha/(3-\alpha)}.
\end{equation}

\section{Covariant holography in a k-essential dark energy universe}\label{sec:covkessenceDE}

In this section we shall follow the covariant formalism developed in \cite{Bousso:2002ju} for general spacetimes in order to study the holographic properties of the spacetime presented in Sec.\ \ref{sec:kessenceDE}.  First, we plot the Penrose diagram for our k-essential asymptotic spacetime. Second, we construct the embedded holographic hypersurfaces (screens) on which the entire bulk information can be stored at a density of no more than one bit per Planck area \cite{'tHooft:1993gx,Susskind:1994vu}. To construct screens, we
slice the spacetime into a family of null hypersurfaces centred at $r=0$ that can be parametrised by time. We end up with two inequivalent null projections, along
past or future-directed light cones, and identify in which direction to project.

In order to determine the causal structure of a homogeneous and isotropic
universe, we conformally map it on a part of the Einstein static universe \cite{Hawking:1973uf} whose causal structure is
that of an infinite cylinder $R \times S^{3}$. The part of it which describes the causal structure of some
arbitrary homogeneous and isotropic universe is bounded by the images of the singularities
and/or past and future causal boundaries. In the resulting Penrose diagram, which represents the conformal structure of infinity of our FRW spacetime, every point represents a $S^{2}$ sphere and each diagonal line represents a light-cone. The two inequivalent null slicings can be represented by the ascending a descending families of diagonal lines.
We then identify the apparent horizons, which are the hypersurfaces on which the expansion of the past or future light-cones vanishes. These horizons are the boundary between a trapped or anti-trapped region and a normal region \cite{Bousso:2002ju,Hawking:1973uf}.

\begin{figure}
\begin{center}
\includegraphics[width=.5\columnwidth]{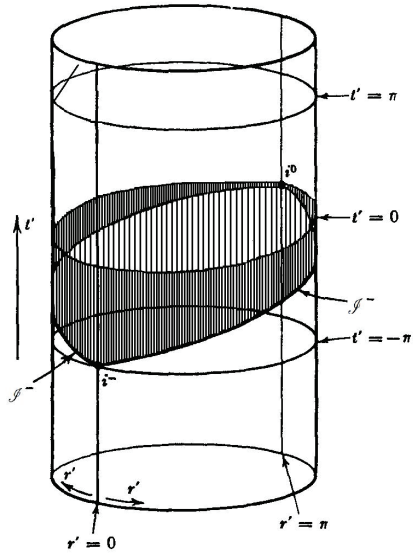}
\caption{\label{fig:2} The flat FRW spacetime filled with a
k-essential scalar field is conformal to the Einstein static
universe for the EoS range $-1<w<-1/3$. This depiction looks similar to
that of the de Sitter space, although it covers a
larger $t'$-interval.}
\end{center}
\end{figure}
We shall finally determine the preferred and
optimal (if any) screen hypersurfaces which are going to encode
all the information in the universe. A preferred screen
is a surface in which the expansion of all
projected null hypersurfaces becomes zero at every point \cite{Bousso:2002ju}. If the expansions of
both independent pairs of orthogonal families of light-rays vanish
on one of the preferred screens, it becomes an optimal screen
\cite{Bousso:2002ju}.

\begin{figure}
\begin{center}
\includegraphics[width=.7
\columnwidth]{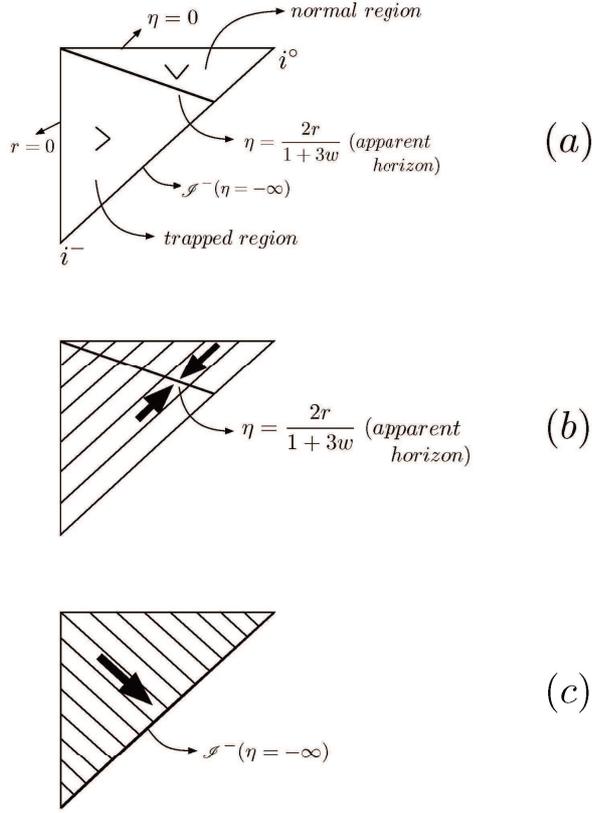}
\caption{\label{fig:3} Penrose diagram of a flat FRW universe dominated by k-essence dark energy for the range $-1<w<-1/3$. The apparent horizon,
$\eta=2r/(1+3w)$, divides the spacetime into a normal
and a trapped region (a). The information contained in the universe can be projected
along future light-cones onto the apparent horizon (b), or
along past light-cones onto past null infinity $\mathcal{I^-}$ (c). Both are preferred screen-hypersurfaces.}
\end{center}
\end{figure}
The metric of a flat FRW spacetime in terms of the conformal time $\eta$ is given by

\begin{equation}\label{confmetric}
ds^2= a(\eta)^2\left(-d\eta^2 +dr^2 +r^2 d\Omega_2^2\right)\;,
\end{equation} where $0 < r < \infty$, and $d\Omega_2^2 = d\theta^2 +\sin^2\theta
d\phi^2$ is the metric on the unit $S^{2}$ sphere, with $0<\theta
<\pi$ and $0 <\phi <2\pi$. This metric can be reduced to a more convenient form \cite{Hawking:1973uf} by defining
some new coordinates, $p$ and $q$, such that
$t'=p+q$ and $r'=p-q$. This way the metric (\ref{confmetric}) can be expressed in a form
which is conformal to that of Minkowski space in spherical
coordinates, and hence locally identical to that of the Einstein static universe

\begin{equation}\label{Einsteinst}
ds^2 =\frac{1}{4}a^2\sec^2 \left[\frac{1}{2}(t'+r')\right]\sec^2
\left[\frac{1}{2}(t'-r')\right]
\left[-(dt')^2+(dr')^2 +\sin^2 r' d\Omega_2^2\right]\;,
\end{equation} where $-\pi <t'+r' <\pi$, $-\pi <t'-r' <\pi$, $r'\geq 0$. The new
coordinates $r'$ and $t'$ are related to the original coordinates
 $\eta$ and $r$ by
\begin{equation}\label{eta}
\eta=\frac{1}{2}\tan\left[\frac{1}{2}(t'+r')\right]+
\frac{1}{2}\tan\left[\frac{1}{2}(t'-r')\right]
\end{equation}
\begin{equation}\label{r}
r=\frac{1}{2}\tan\left[\frac{1}{2}(t'+r')\right]-
\frac{1}{2}\tan\left[\frac{1}{2}(t'-r')\right]\;.
\end{equation}
Now our flat FRW spacetime filled with a k-essence scalar field and
whose EoS  lies in the range $-1<w<-1/3$
can be mapped into the part of
the Einstein static universe determined by the values taken
by $\eta$ in the interval $-\infty<\eta<0$, which corresponds to the ranges $-\pi<t'<0$ and $0 < r' <\pi$. From these, the resulting Penrose diagram follows.

 In Fig.\ \ref{fig:2}, we show the parts of the Einstein
static cylinder which are conformal to the k-essential flat FRW
spacetime for $-1<w<-1/3$ . The conformal region runs from $t'=0$ to an extreme $t'< 0$. The corresponding Penrose diagram is plotted in Fig.\ \ref{fig:3}.

Finally, following the procedure devised in \cite{Bousso:2002ju}, we are in a position to
construct the holographic hypersurfaces (screens)  in a flat FRW universe dominated by k-essence dark energy.
The apparent horizon is given by
$\eta=2r/(1+3w)$. The interior of the apparent horizon,  $\eta\geq 2r/(1+3w)$, can be projected along future light-cones centred at $r=0$, or by
space-like projection, onto the apparent horizon.
 The exterior, $\eta\leq 2r/(1+3w)$, can also be
projected by future light cones, but in the opposite direction, onto the apparent horizon. Otherwise, the entire flat k-essential dark energy universe can be projected along past light-cones onto the past null
infinity. The two holographic preferred screen hypersurfaces, given by the apparent horizon
$\eta=2r/(1+3w)$ and the past null infinity $\mathcal{I}^-$, are depicted in Fig.\ \ref{fig:3}.

Notice that $ln(a_{0}-t_{0}) < \eta < \infty$ when $w=-1/3$. In this case there is no event horizon (although an infinitesimal acceleration would raise it) but there is an apparent one located at $\eta = \infty$.

\section{Covariant holography in a k-essential phantom dark energy universe}\label{sec:phkessenceDE}

Phantom dark energy \cite{Caldwell:1999ew,Starobinsky:1999yw} is a well established dark energy candidate \cite{Carroll:2003st,Singh:2003vx,Cline:2003gs,Sami:2003xv}. Planck latest results \cite{Ade:2013zuv} plus WMAP low-\emph{l} polarisation (WP), when combined with Supernova Legacy Survey (SNLS) data, favour the phantom domain at 2$\sigma$ level for a constant $w$
\begin{equation}\label{Planck}
w=- 1.13^{+0.13}_{-0.14} \;(95\%; Planck+WP+SNLS)\;,
\end{equation} while the Union2.1 compilation of 580 Type Ia supernovae (SNe Ia) is more consistent
with a cosmological constant ($w=-1$). If we combine Planck+WP with measurements of $H_{0}$ \cite{Riess:2011yx}, we get for a constant $w$
\begin{equation}\label{Planck2}
w=- 1.24^{+0.18}_{-0.19}\;
\end{equation} which is in tension with $w=-1$ at more than the 2$\sigma$ level. Moreover, claims for $w<-1$ at $\geq 2\sigma$ have
been presented, such as \cite{Rest:2013bya}, which features high-quality
data and a careful analysis including systematic errors
\cite{Scolnic:2013aya}. Also, the authors in \cite{Shafer:2013pxa} found that for the SNLS3 and the Pan-STARRS1 survey (PS1 SN)
data sets, the combined SNe Ia + Baryon Acoustic Oscillations (BAO) + Planck data
yield a phantom EoS at $\sim 1.9\sigma$ confidence.

For the phantom case we can express $g(y)$ as \cite{GonzalezDiaz:2003rf},
\begin{equation}\label{gphantom}
g(y)=By^{\beta}\;,
\end{equation} where $B$ and $\beta$ are given constants such that $B>0$ and $\beta \left(=-\frac{1}{w}\right)$ satisfies the observational constraint $0<\beta<0.9$ \cite{Ade:2013zuv}.

From Eq. (\ref{sofs}) we see that $c_{s}^{2}<0$ in this case. This will give rise to instabilities at the level of perturbations. However, since in this work we are only interested in the background evolution of the universe, we can overlook this problem.

\begin{figure}
\begin{center}
\includegraphics[width=.5\columnwidth]{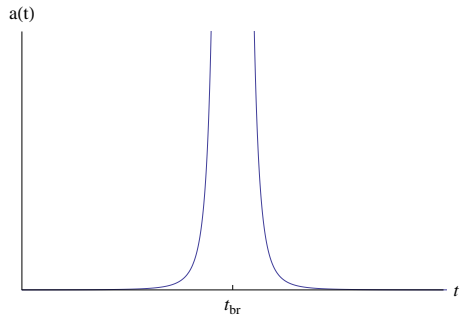}
\caption{\label{fig:4} Generic shape of the evolution of the scale factor $a(t)$ with cosmological time $t$ for the function $g(y)$ given in Eq. (\ref{gphantom}). All units in the plot are arbitrary.}
\end{center}
\end{figure}

Through Eq. (\ref{gphantom}) we obtain the scale factor
\begin{equation}\label{scalefactorphantom}
a(t)=a_{0}(t-t_{br})^{-2\beta/[3(1-\beta)]}\;,
\end{equation} where $a_{0}$ is an arbitrary initial value for the scale factor and $t_{br}$ is the time at the big rip, which is also arbitrary in this model. We have shown the evolution of the scale factor with respect to cosmic time $t$ in Fig.\ \ref{fig:4}.

As we did in Sect.\ \ref{sec:kessenceDE} , we use the conformal time for convenience

\begin{equation}\label{confphantom}
\eta=\int\frac{dt}{a(t)} = \frac{3(1-\beta)}{a_{0}(3-\beta)}(t-t_{br})^{(3-\beta)/[3(1-\beta)]}\;,
\end{equation}
so that the scale factor (\ref{scalefactorphantom}) becomes
\begin{equation}\label{scphconf}
a(\eta)=a_{0}^{3(1-\beta)/[3-\beta]}\left[\frac{(3-\beta)\eta}{3(1-\beta)}\right]^{-2\beta/(3-\beta)}\;.
\end{equation}
Armed with Eq.\ (\ref{scphconf}) we can again use the metric of a flat
FRW spacetime Eq.\ (\ref{confmetric}). Since the holographic preferred screens will always be described in the causal development as the latest surface which is expanding, we must allow the causal evolution to reach at least a little region beyond that holographic surface, into a well-defined contracting region. Now,
in order for the time coordinate $t$ to run beyond
the big rip barrier located at $t=t_{br}$, not all the values
of the parameter $\beta$ are allowed but only those that do not lead to negative or
imaginary values for the scale factor $a(t)$. In particular, only the values for $\beta$ satisfying the following relation \cite{GonzalezDiaz:2005sh}
\begin{equation}\label{discretebeta}
\beta=\frac{3n}{3n+1},\;\;\; n=1,2,3,...
\end{equation}
where $n$ is necessarily finite as $\beta<1$, will guarantee real and positive values for $a(t)$ from $t=0$ to
$t=\infty$. If we insert now Eq.\ (\ref{discretebeta}) into
Eqs.\ (\ref{confphantom}) and (\ref{scphconf}), we arrive at
\begin{equation}\label{conformalphdisc}
\eta=\frac{(t-t_{br})^{2n+1}}{a_{0}(2n+1)}
\end{equation} and
\begin{equation}
a(\eta)=a_{0}^{1/(2n-1)}\left[(2n+1)\eta\right]^{-2n/(2n+1)}\;.
\end{equation}

\begin{figure}
\begin{center}
\includegraphics[width=.5\columnwidth]{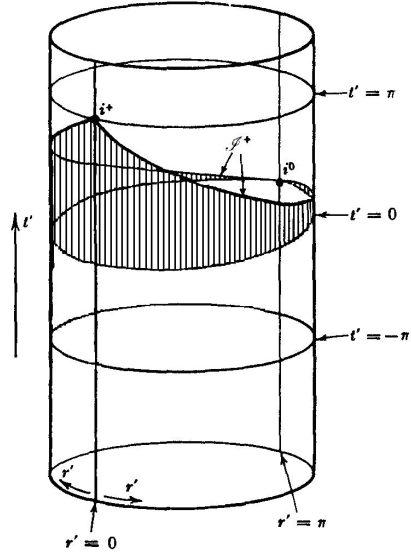}
\caption{\label{fig:5} The flat FRW spacetime filled with a
k-essential phantom energy scalar field is conformal to the Einstein static
universe for the range $-\infty<w<-1$.}
\end{center}
\end{figure}
Proceeding now as in Sect.\ \ref{sec:kessenceDE}, we locally obtain the metric (\ref{Einsteinst}) for the Einstein
static universe, where $\eta$ and $r$ given by Eqs.\ (\ref{eta}) and
(\ref{r}). By inspecting Eq.\ (\ref{conformalphdisc}) we see that in the phantom scenario $\eta$ should
run within the interval starting from
$\eta= \frac{3(1-\beta)}{a_{0}(3-\beta)}(t_{0}-t_{br})^{(3-\beta)/[3(1-\beta)]}\equiv\eta_0 < 0$
for $t=t_0$, reaching $\eta=0$ at the big rip
singularity ($t=t_{br}$), to finally get to
positive infinity as $t\rightarrow\infty$; i.e. $\eta_0 <\eta<+\infty$.
\begin{figure}
\begin{center}
\includegraphics[width=.95\columnwidth]{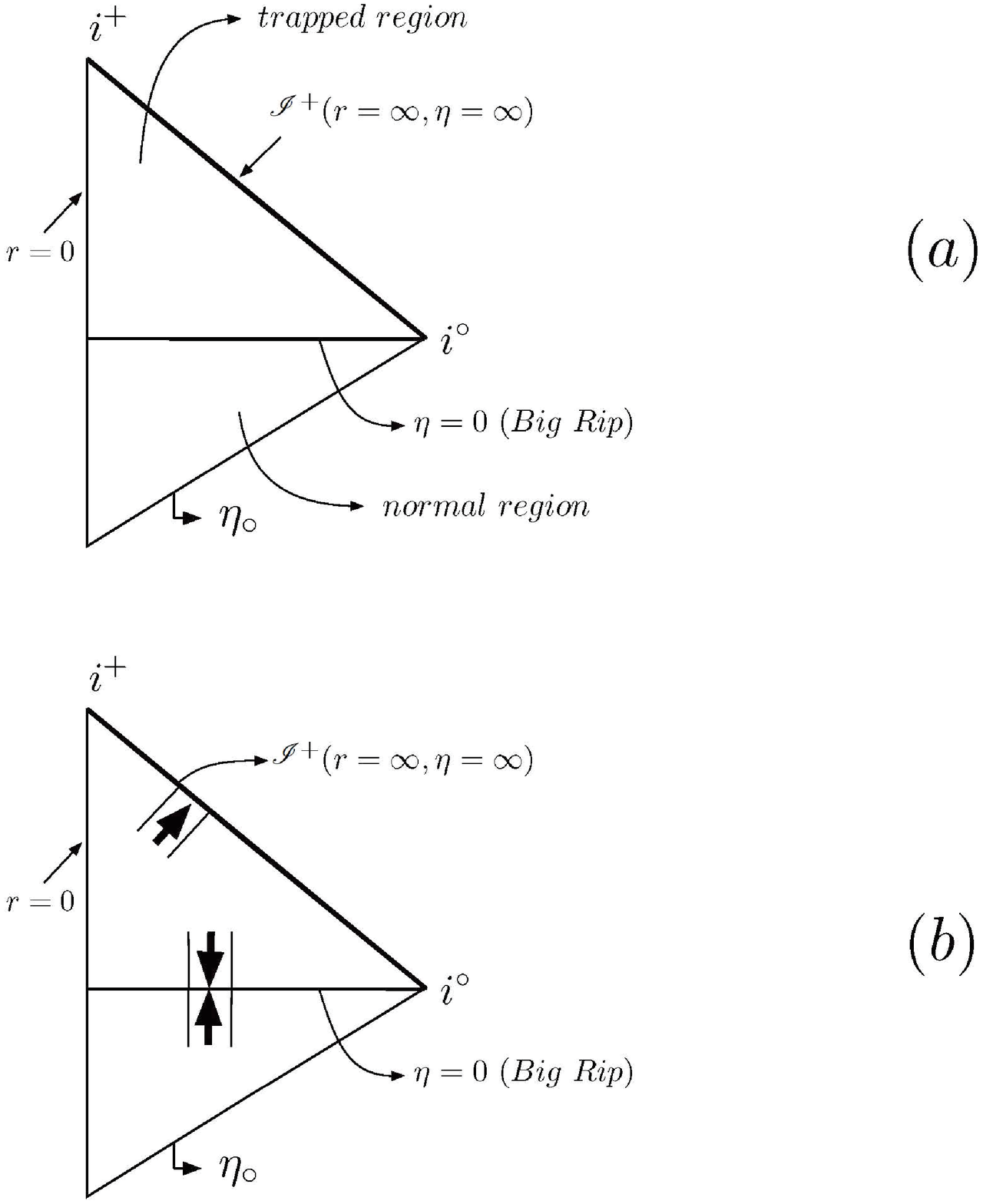}
\caption{\label{fig:6} Penrose diagram of a flat  FRW
universe filled with a k-essence phantom scalar field for
$-\infty<w<-1$. The apparent horizon, also located at
$\eta=2r/(1+3w)$, divides the spacetime into a normal
and a trapped region (a). The information contained in the universe can be projected
along future light-cones from the normal region or
along past light-cones from the trapped region, both onto the apparent big rip horizon. It can also be projected along future light cones onto the future null infinity $\mathcal{I^+}$ (b). Both the big rip and $\mathcal{I^+}$ are preferred screen-hypersurfaces.}
\end{center}
\end{figure}
Therefore, the flat spacetime filled with a phantom k-essence field above considered can also be
mapped into those parts of the cylindric Einstein static universe
which are determined by the values of
conformal time $\eta$ we have just discussed. The part $0<\eta<+\infty$
of the whole interval $\eta_0 <\eta< +\infty$ will correspond to
the range $0<t'<\pi$ and $0<r'<\pi$, and the part $\eta_0 <\eta<0$
will correspond to a $\beta$-dependent subinterval of the range
 $-\pi<t'<0$ and $0<r'<\pi$. This mapping is shown in Fig.\ \ref{fig:5}
and the corresponding Penrose diagram is depicted in Fig.\ \ref{fig:6}.

\section{Holographic dark energy models}\label{sec:HDE}
Based on the holographic bound on the entropy \cite{GonzalezDiaz:1983yf,'tHooft:1993gx,Susskind:1994vu} and on the validity of effective
local quantum field theory in a box of size L, Cohen et al
\cite{Cohen:1998zx} suggested a relationship between the ultraviolet
and the infrared cutoffs due to the limit set by the
formation of a black hole. This led Li to propose
a HDE model \cite{Li:2004rb} that can explain the accelerated expansion of the universe. There are several incarnations of the HDE model depending on the choice for the infrared cutoff \cite{Li:2011sd} but the most widely studied \cite{Li:2004rb} incorporates the future event horizon as the infrared cutoff and holographic screen.
This model complies well with observational data \cite{Huang:2004wt,Zhang:2005hs,Chang:2005ph,Ma:2007pd,Li:2009bn,Li:2013dha,Xu:2013mic}, but has attracted some criticisms, known as the causality and circularity problems \cite{Kim:2012ik}. However, Li has recently proposed a new HDE model with action principle \cite{Li:2012xf} which is apparently free from these problems and in which the future event horizon is not an input but automatically follows from the equations of motion. This new HDE model also provides a nice fit to the most recent cosmological data \cite{Li:2012fj}. In a flat universe dominated by dark energy, the Li model \cite{Li:2004rb} predicts the following relation between
the Hubble parameter and the size of the future event horizon
$R_h$
\begin{equation}\label{Li}
H^2=\frac{\dot{a}^{2}}{a^{2}}=\frac{8\pi G\rho_{\phi}}{3}=\frac{c^2}{R_h^2}\;,
\end{equation}
where $R_h=a(t)\int_t
^{\infty}dt'/a(t')$ is the proper size of the future event horizon which plays the role of the holographic screen and $c$ is a numerical parameter of order unity which
is related to $\alpha$ by $\alpha=3/(1+2/c)$ and to $w$ by $w=-(1+2/c)/3$.
For our model in Sec.\ \ref{sec:kessenceDE}, with the scale factor given by Eq.
(\ref{scalefactor}), we get for the size of the future event horizon

\begin{eqnarray}
R_h & = & \frac{2\alpha}{\alpha-3}\left[a_{0}^{-\frac{3(1-\alpha)}{2\alpha}}-\frac{3}{2}\left(\frac{1-\alpha}{\alpha}\right)(t-t_{0})\right]^{-2\alpha/[3(1-\alpha)]}\nonumber\\
& &\times\left.\left\{\left[a_{0}^{-\frac{3(1-\alpha)}{2\alpha}}-\frac{3}{2}\left(\frac{1-\alpha}{\alpha}\right)(t-t_{0})\right]^{-(3-\alpha)/2\alpha}\right\}\right|^{\infty}_{t}\;,
\end{eqnarray} which is finite for finite t since $c > 1$ (which implies that $w>-1$).
Therefore, in this case Eq. (\ref{Li}) holds and
hence the Li model is well defined.

For the phantom case of Sec.\ \ref{sec:phkessenceDE},  $c<1$ (i.e. $w<-1$), and the proper size of the future event horizon becomes
infinity, or equivalently, will vanish for phantom energy
\begin{equation}
R_h  =  \frac{3(1-\beta)\left(t-t_{br}\right)^{-2\beta/[3(1-\beta)]}}{3-\beta}
\left.\left[\left(t'-t_{br}\right)^{(3-\beta)/[3(1-\beta)]}\right]\right|^{\infty}_{t-t_{br}}=\infty\;.
\end{equation} This means that Eq. (\ref{Li}) is ill-defined for $c<1$ since it leads to $H=0$, and in this phantom case we should use instead \cite{GonzalezDiaz:2005sh}

\begin{equation}
H_{{\rm ph}}^2 =\frac{\dot{a}^2}{a^2}=\frac{8\pi
G\rho_{\phi}}{3}=\frac{c^2}{R_{br}^2}\;,
\end{equation} where
\begin{equation}\label{Rbr1}
R_{br}=a(t)\int_t^{t_{br}}\frac{dt'}{a(t')}=\frac{t_{br}-t}{2n+1}
\end{equation}
is the proper size of the future event horizon for the k-essential phantom model. Given that Eq. (\ref{Li}) is no longer valid for a covariant description of
an accelerating cosmic holography, the two holographic screens that correspond to the covariant treatment are the one
at the big rip hypersurface and the one at the future null infinity $\mathcal{I^+}$ as found in Sec.\ \ref{sec:phkessenceDE}.

\section{Conclusions}\label{sec:concl}

We have studied the holography of an accelerating
universe within the framework of a flat FRW cosmology when the dark energy is due to the presence of a k-essence scalar field and for any constant EoS $w$. We have done so by applying
a covariant formalism \cite{Bousso:2002ju} and comparing the results with the well-known HDE model \cite{Li:2004rb,Li:2012xf}.

We have found that the covariant formalism leads to two different
holographic preferred screens. For the $w>-1$ case, one is located at the apparent
horizon at $\eta=2r/(1+3w)$ and the other one at the past null
infinity $\mathcal{I^-}$. On the other hand, for $w<-1$, one preferred screen is located at the apparent
horizon at $\eta=2r/(1+3w)$, which in this case is the big rip hypersurface, and the other one at the future null infinity $\mathcal{I^+}$. These results, as opposed to the ones obtained by using the HDE model \cite{Li:2004rb,Li:2012xf} whose holographic screen is positioned at the future event horizon, allow the development of fundamental theories such as string theory. In particular, the construction of a conventional S-matrix should be possible when one comes near $\mathcal{I^-}$ or $\mathcal{I^+}$.

A seemingly problematic issue that we may encounter is the possible contradiction that might exist between the implications from the covariant treatment of phantom holography and the result that phantom energy is characterised by a negative temperature \cite{GonzalezDiaz:2004eu}. It could well be thought, at first sight,  that if the preferred holographic screens for phantom energy are placed at the big rip and the future null infinity, then the entropy that should be associated with that phantom fluid would be negative definite, leading therefore to a definite positive temperature. However, this line of reasoning is flawed because the entropy that we should consider for this kind of analysis is the one defined by the surface area of the future preferred holographic screen, which in this case turns out to be given by (see Eq.\ (\ref{Rbr1}))

\begin{equation}\label{Rbr2}
R_{br}=\frac{-3(1-\beta)(t-t_{br})}{(3-\beta)}
\end{equation}

The relevant entropy here would coincide with the entropy of entanglement \cite{Muller:1995mz} and would be given by

\begin{equation}\label{ententropy}
S_{Ent}=\gamma R_{br}^{2}=\gamma \left[\frac{3(1-\beta)(t-t_{br})}{(3-\beta)}\right]^{2}|_{t>t_{br}}
\end{equation} with $\gamma$ being a constant of order unity and where in order to calculate the entropy of entanglement we have used the equivalence between the regions before and after the big rip hypersurface. In this case, we have integrated out the region before that surface. This entanglement entropy is definite positive and increases with time, leading again to the conclusion that the temperature of a phantom fluid is definite negative.

Note that the existence of an entanglement entropy \cite{Muller:1995mz} associated to correlations across the finite future horizons signals a close connection between the entropy of entanglement and the covariant definition of holography \cite{Bousso:2002ju}.

\begin{acknowledgements}
The author is grateful to Andrea Maselli for help with the plots and to Diego Pav\'on for reading the manuscript. This work was supported by the 'Fundaci\'on Ram\'on Areces' and Ministerio de Econom\'ia y Competitividad (Spain) through project number FIS2012-38816.
\end{acknowledgements}

\bibliographystyle{spphys}       
\bibliography{KCHGERG}   


\end{document}